\documentclass[11pt]{siamltex}%
\usepackage{amsmath}
\usepackage{amsfonts}
\usepackage{amssymb}
\usepackage{graphicx}%
\usepackage{natbib}
\usepackage{fullpage}
\providecommand{\U}[1]{\protect\rule{.1in}{.1in}}
\newcommand{\unit}[1]{\operatorname{#1}}
\renewcommand{\vec}[1]{\boldsymbol{#1}}
\def\refname{\centering \sc References}
\begin{document}
%
%TCIMACRO{\TeXButton{Front matter}{\linenumbers\title
%{Spectral Diagonal Ensemble Kalman Filters}
%\Author[1]{Ivan}{Kasanick\'{y}}
%\Author[1,2]{Jan}{Mandel}
%\Author[1]{Martin}{Vejmelka}
%\affil
%[1]{Institute of Computer Science, Academy of Sciences of the Czech Republic}
%\affil
%[2]{Department of Mathematical and Statistical Sciences, University of Colorado Denver}
%\runningtitle{Spectral Diagonal Ensemble Kalman Filters}
%\runningauthor{Ivan Kasanick\'{y}, Jan Mandel, and Martin Vejmelka}
%\correspondence{Jan Mandel (jan.mandel@gmail.com)}
%\received{}
%\pubdiscuss{}
%\revised{}
%\accepted{}
%\published{}
%\firstpage{1}
%\maketitle\begin{abstract}
%A new type of ensemble Kalman filter is developed, which is based on replacing the sample covariance in the
%analysis step by its diagonal in a spectral basis. It is proved that this technique
%improves the aproximation of the covariance when the covariance itself is
%diagonal in the spectral basis, as is the case, e.g., for a second-order
%stationary random field and the Fourier basis. The method is extended by wavelets to the case when the
%state variables are random fields which are not spatially homogeneous. Efficent implementations by the
%fast Fourier transform (FFT) and
%discrete wavelet transform (DWT) are presented for several types of observations, including high-dimensional data given on a part of the domain.
%Computational experiments confirm that the method performs
%well on the Lorenz 96 problem and the shallow water equations with very small ensembles and over multiple analysis cycles..
%\end{abstract}}}%
%BeginExpansion
\title{Spectral Diagonal Ensemble Kalman Filters}
\author{{Ivan} {Kasanick\'{y}}\footnotemark[1] 
\and {Jan} {Mandel}\footnotemark[1]~\footnotemark[2]
\and {Martin} {Vejmelka}\footnotemark[1] 
}
\footnotetext[1]{Institute of Computer Science, Academy of Sciences of the Czech Republic}
\footnotetext[2]{Department of Mathematical and Statistical Sciences, University of Colorado Denver}
%\runningtitle{Spectral Diagonal Ensemble Kalman Filters}
%\runningauthor{Ivan Kasanick\'{y}, Jan Mandel, and Martin Vejmelka}
%\correspondence{Jan Mandel (jan.mandel@gmail.com)}
%\received{}
%\pubdiscuss{}
%\revised{}
%\accepted{}
%\published{}
%\firstpage{1}
\maketitle\begin{abstract}
A new type of ensemble Kalman filter is developed, which is based on replacing the sample covariance in the
analysis step by its diagonal in a spectral basis. It is proved that this technique
improves the aproximation of the covariance when the covariance itself is
diagonal in the spectral basis, as is the case, e.g., for a second-order
stationary random field and the Fourier basis. The method is extended by wavelets to the case when the
state variables are random fields which are not spatially homogeneous. Efficient implementations by the
fast Fourier transform (FFT) and
discrete wavelet transform (DWT) are presented for several types of observations, including high-dimensional data given on a part of the domain, such as radar and satellite images.
Computational experiments confirm that the method performs
well on the Lorenz 96 problem and the shallow water equations with very small ensembles and over multiple analysis cycles..
\end{abstract}%
%EndExpansion

%\introduction
\section{Introduction}

Data assimilation consists of incorporating new data periodically into
computations in progress, which is of interest in many fields, including
weather forecasting \citep[e.g.,][]{Kalnay-2003-AMD,Lahoz-2010-DAM}. One data
assimilation method is filtering \citep[e.g.,][]{Anderson-1979-OF}, which is a
sequential Bayesian estimation of the state at a given time given the data
received up to that time. The probability distribution of the system state is
advanced in time by a computational model, while the data is assimilated by
modifying the probability distribution of the state by an application the
Bayes theorem, called analysis. In the methods considered here, data is
assimilated in discrete time steps, called analysis cycles, and the
probability distributions are represented by their mean and covariance (thus
making a tacit assumption that they are at least close to gaussian). When the
state covariance is given externally, bayesian estimation becomes the
classical optimal statistical interpolation (OSI). The Kalman filter (KF) uses
the same computation as OSI in the analysis, but it evolves the covariance matrix
of the state in time along with the model state. Since the covariance matrix
can be large, the KF is not suitable for high-dimensional systems. The
ensemble Kalman filter (EnKF) \citep{Evensen-2009-DAE} replaces the state
covariance by the sample covariance computed from an ensemble of simulations,
which represent the state probability distribution. It can be proved that the
EnKF\ converges to the KF in the large ensemble limit
\citep{Kwiatkowski-2014-CSR,LeGland-2011-LSA,Mandel-2011-CEK} in the gaussian
case, but an acceptable approximation may require hundreds of ensemble members
\citep{Evensen-2009-DAE}, because of spurious long-distance correlations in
the sample covariance due to its low rank. Localization techniques
\citep[e.g.,][]{Anderson-2001-EAK,Furrer-2007-EHP,Hunt-2007-EDA}, essentially
suppress long-distance covariance terms \citep{Sakov-2010-RBT}, which improves
EnKF performance for small ensembles.

FFT EnKF \citep{Mandel-2010-DDC,Mandel-2010-FFT} was proposed as an
alternative approach to localization, based on replacing the sample covariance
in the EnKF by its diagonal in the Fourier space. This approach is motivated
by the fact that a random field in cartesian geometry is second order
stationary (that is, the covariance between the values at two points depends
only on their distance vector) if and only if its covariance in the Fourier
space is diagonal \citep[e.g.,][]{Pannekoucke-2007-FPW}. On a sphere, an
isotropic random field has diagonal covariance in the basis of spherical
harmonics \citep{Boer-1983-HIT}, so similar algorithms can be developed there
as well. However, the stationarity assumption does not allow the covariance to
vary spatially. For this reason, the FFT EnKF was extended to wavelet EnKF
\citep{Beezley-2011-WEK}. The use of wavelets results in an automatic
localization, which varies in space adaptively. For wavelets, the effect of
the diagonal spectral approximation is equivalent to a weighted spatial
averaging of local covariance functions \citep{Pannekoucke-2007-FPW}. Diagonal
matrices are cheap to manipulate computationally, but implementing the
multivariate case and general observation functions is not straighthforward.

Diagonal spectral approximation and, more generally, sparse spectral
approximation, have been used as a statistical model for the background
covariance in data assimilation in meteorology for some time. The optimal
statistical interpolation system from \citet{Parrish-1992-NMC} was based on a
diagonal approximation in spherical harmonics, already used as horizontal
basis functions in the model, and a change of state variables into physically
balanced analysis variables. The ECMWF\ 3DVAR system \citep{Courtier-1998-EIT}
also used diagonal covariance in spherical harmonics. Diagonal approximation
in the Fourier space for homogeneous 2D error fields, with physically balanced
crosscovariances, was proposed in \cite{Berre-2000-ESM}. The Fourier
diagonalization approach was extended by \cite{Pannekoucke-2007-FPW} to sparse
representation of the background covariance by thresholding wavelet
coefficients, and into a combined spatial and spectral localization by
\cite{Buehner-2007-SSL}.

While modeling of background covariances typically uses multiple sources
including historical data, the EnKF builds the covariance in every analysis
cycle from the ensemble itself. In this paper, we prove that replacing the
sample covariance by its spectral diagonal improves the approximation when the
covariance itself is diagonal in the spectral space, as is the case, e.g.,
when the state is a second order stationary random field and a Fourier basis
is used. The result, however, is general and it applies to an arbitrary
orthogonal basis, including wavelets. We also develop computationally
efficient spectral EnKF\ algorithms, which take advantage of the diagonal form
of the covariance, in the multivariate case and for several important classes
of observations. We demonstrate the methods on computational examples with the
Lorenz 96 system and shallow water equations, which show that good performance
can be achieved with very small ensembles.

\section{Notation}

Vectors in $\mathbb{R}^{n}$ or $\mathbb{C}^{n}$ are typeset as $\vec{u}$ and
understood to be columns. Random vectors are typeset as $\vec{X}$. The entry
$i$ of $\vec{X}$ is denoted by $\left(  \vec{X}\right)  _{i}$. Matrices
(random or deterministic) are typeset as $\mathbf{A}$, and and $\mathbf{A}%
^{\ast}$ is the transpose, or conjugate transpose in the complex case. The
entry $i,j$ of matrix $\mathbf{A}$ is denoted by $\left(  \mathbf{A}\right)
_{i,j}$ or $a_{i,j}$, and $\mathbf{A}=\left[  \vec{a}_{1},\ldots,\vec{a}%
_{n}\right]  $ is the writing of a matrix as a collection of columns.
Nonlinear operators are typeset as $\mathcal{M}$. The mean value is denoted by
$\operatorname{E}\left[  \cdot\right]  $, and $\operatorname{Var}$ is the
variance. $N\left(  0,1\right)  $ is the normal (gaussian) distribution with
zero mean and unit variance, and $N\left(  \vec{m},\mathbf{C}\right)  $ is the
multivariate normal distribution with mean $\vec{m}$ and covariance
$\mathbf{C}$. The Euclidean norm of a vector is $\left\Vert \vec{u}\right\Vert
=\left(  \sum_{i=1}^{n}\left\vert u_{i}\right\vert ^{2}\right)  ^{1/2}$. The
Frobenius norm of a matrix is $\Vert\mathbf{A}\Vert_{\mathrm{F}}=\left(
\sum_{i=1}^{m}\sum_{j=1}^{n}|a_{i,j}|^{2}\right)  ^{1/2}.$

\section{Kalman filter and ensemble Kalman filter}

The state of the system at time $t$ is described by a random vector $\vec
{X}_{t}$ of length $n$. The system evolution between two times $t_{1}$ and
$t_{2}$ is given by a function $\mathcal{M}(.,t_{1},t_{2})$, so that
\begin{equation}
\vec{X}_{t_{2}}^{\mathrm{f}}=\mathcal{M}(\vec{X}_{t_{1}}^{\mathrm{a}}%
,t_{1},t_{2}). \label{eq:kfmodel}%
\end{equation}
The goal of the Kalman filter (KF) \citep{Kalman-1960-NAL} is to correct the
forecast state of the system $\vec{X}_{t}^{\mathrm{f}}$ to obtain the analysis
estimate $\vec{X}_{t}^{\mathrm{a}}$ of the true state $\vec{X}_{t}$, given
noisy observations $\vec{Y}_{t}=\mathbf{H}_{t}\vec{X}_{t}+\mathbf{\epsilon
}_{t}$, where $\mathbf{H}_{t}$ is an observation operator, i.e., a mapping
from state space to a data space, and $\epsilon_{t}\sim N\left(  \vec
{0},\mathbf{R}_{t}\right)  $. When the distributions of the state $\vec{X}%
_{t}$ and the data error are gaussian, the analysis satisfies
\begin{equation}
\vec{X}_{t}^{\mathrm{a}}=\vec{X}_{t}^{\mathrm{f}}-\mathbf{C}_{t}\mathbf{H}%
_{t}^{\ast}\left(  \mathbf{H}_{t}\mathbf{C}_{t}\mathbf{H}_{t}^{\ast
}+\mathbf{R}_{t}\right)  ^{-1}\left(  \mathbf{H}_{t}\vec{X}_{t}^{\mathrm{f}%
}-\vec{Y}_{t}\right)  , \label{eq:kfupdate}%
\end{equation}
where $\mathbf{C}_{t}$ is the covariance of the forecast $\vec{X}%
_{t}^{\mathrm{f}}$. In the KF, the state is represented by its mean and
covariance, and the mean is transformed also by (\ref{eq:kfmodel}) and
(\ref{eq:kfupdate}). In the rest of the paper, we will drop the time index $t$
and the superscript $\mathrm{f}$, unless there is a danger of confusion.

In the EnKF, the analysis formulas (\ref{eq:kfmodel}) and (\ref{eq:kfupdate})
are applied to each ensemble member, with the covariance replaced by the
sample covariance from the ensemble. The resulting ensemble, however, would
underestimate the analysis covariance, which is corrected by a data
perturbation by sampling from the data error distribution
\citep{Burgers-1998-ASE}. Denote by $\vec{X}^{1},\ldots,\vec{X}^{N}$ the
forecast ensemble, created either by a perturbation of a background state or
by evolving each analysis ensemble member from the previous time step
independently by (\ref{eq:kfmodel}). Then, the analysis ensemble members are%
\begin{equation}
\vec{X}^{\mathrm{a},j}=\vec{X}^{j}-\mathbf{C}^{N}\mathbf{H}^{\ast}\left(
\mathbf{HC}^{N}\mathbf{H}^{\ast}+\mathbf{R}\right)  ^{-1}\left(
\mathbf{H}\vec{X}^{j}-\vec{Y}^{j}\right)  , \label{eq:enkfupdate}%
\end{equation}
where the sample covariance matrix is%
\begin{equation}
\mathbf{C}^{N}=\frac{1}{N-1}\sum_{j=1}^{N}\left(  \vec{X}^{j}-\bar
{\vec{X}}\right)  \left(  \vec{X}^{j}-\bar{\vec{X}}\right)  ^{\ast}%
,\quad\bar{\vec{X}}=\frac{1}{N}\sum_{j=1}^{N}\vec{X}^{j}
\label{eq:samplecov}%
\end{equation}
and $\vec{Y}^{j}=\vec{Y}+\vec{\tau}^{j}$ are the perturbed observations, with
$\vec{\tau}^{j}\sim N\left(  0,\mathbf{R}\right)  $ independent.

The advantage of the EnKF update formula (\ref{eq:kfupdate}) is that it can be
implemented efficiently without having acces to the whole sample covariance
matrix $\mathbf{C}^{N}$. On the other hand, the rank of matrix $\mathbf{C}%
^{N}$ is at most $N-1$, and, in the usual case when $N<<n$, the low rank of
the approximation $\mathbf{C}^{N}$ of the true forecast covariance
$\mathbf{C}$ is the biggest drawback of the EnKF.

\section{Spectral diagonal EnKF}

\label{sec:sdenkf}

Let $\mathbf{F}$ be an orthonormal transformation matrix, which transform each
ensemble member to spectral space, and denote each transformed ensemble member
by the additional subscript $\mathbf{F}$, $\vec{X}_{\mathbf{F}}^{j}%
=\mathbf{F}\vec{X}^{j}$, $j=1,\ldots,N$. Since the transformation is
orthonormal, the inverse transformation is $\mathbf{F}^{\ast}$, so
$\mathbf{F}^{\ast}\vec{X}_{\mathbf{F}}^{j}=\vec{X}^{j}$ for each
$j=1,\ldots,N.$ The columns of the inverse transform matrix $\mathbf{F}^{\ast
}$ are the spectral basis elements $\vec{u}_{1},\ldots,\vec{u}_{n}$, i.e.,
$\mathbf{F}=\left[  \vec{u}_{1},\ldots,\vec{u}_{n}\right]  ^{\ast}$. We will
also denote the sample covariance of the transformed ensemble with the
additional subscript $\mathbf{F}$,
\begin{equation}
\mathbf{C}_{\mathbf{F}}^{N}=\frac{1}{N-1}\sum_{j=1}^{N}\left(  \vec
{X}_{\mathbf{F}}^{j}-\bar{\vec{X}}_{\mathbf{F}}\right)  \left(  \vec
{X}_{\mathbf{F}}^{j}-\bar{\vec{X}}_{\mathbf{F}}\right)  ^{\ast
}=\mathbf{FC}^{N}\mathbf{F}^{\ast},\quad\bar{\vec{X}}_{\mathbf{F}}%
=\frac{1}{N}\sum_{j=1}^{N}\vec{X}_{\mathbf{F}}^{j}.
\label{eq:spectral-sample-cov}%
\end{equation}
The idea of the spectral diagonal Kalman filter is to replace the sample
covariance in the update formula (\ref{eq:enkfupdate}) by only the diagonal
elements of sample covariance in spectral space,%
\begin{equation}
\mathbf{D}_{\mathbf{F}}^{N}=\mathbf{C}_{\mathbf{F}}^{N}\circ\mathbf{I}%
,\quad\left[
\begin{matrix}
c_{1,1} & 0 & \cdots & 0\\
0 & c_{2,2} &  & \vdots\\
\vdots &  & \ddots & 0\\
0 & \cdots & 0 & c_{n,n}%
\end{matrix}
\right]  ,\quad c_{i,i}=\frac{1}{N-1}\sum_{j=1}^{N}\left\vert \bigl(\vec
{X}_{\mathbf{F}}^{j}\bigr)_{i}-\bigl(\bar{\vec{X}}_{\mathbf{F}}%
\bigr)_{i}\right\vert ^{2}. \label{eq:def-D}%
\end{equation}
where $\circ$ stands for Schur product, i.e., element-wise multiplication. The
entries $c_{i,i}$ are the sample variances, computed without forming the whole
matrix $\mathbf{C}_{\mathbf{F}}^{N}$. The diagonal approximation is
transformed back to physical space as
\begin{equation}
\mathbf{D}^{N}=\mathbf{F}^{\ast}\mathbf{D}_{\mathbf{F}}^{N}\mathbf{F,}
\label{eq:sdapprox}%
\end{equation}
and the proposed analysis update is then
\begin{equation}
\vec{X}^{\mathrm{f},j}=\vec{X}^{j}-\mathbf{D}^{N}\mathbf{H}\left(
\mathbf{H}\mathbf{D}^{N}\mathbf{H}^{\ast}+\mathbf{R}\right)  ^{-1}\left(
\mathbf{H}\vec{X}^{j}-\vec{Y}^{j}\right)  . \label{eq:sdenkfupdate}%
\end{equation}

\section{Error analysis}

\label{sec:error-analysis}

We will now compare the expected errors of the sample covariance and its
spectral diagonal approximation (\ref{eq:spectral-sample-cov}). Assume that
the ensemble members $\vec{X}^{i}\sim N\left(  \bar{\vec{X}}%
,\mathbf{C}\right)  $ are independent, and the columns of the inverse spectral
transformation $\mathbf{F}^{\ast}$ are eigenvectors $\vec{u}_{i}$ of the
covariance $\mathbf{C}$ with the corresponding eigenvalues $\lambda_{i}$,%
\begin{equation}
\mathbf{F}=\left[  \vec{u}_{1},\ldots,\vec{u}_{n}\right]  ^{\ast}%
,\quad\mathbf{Cu}_{i}=\lambda_{i}\vec{u}_{i},\quad\mathbf{FF}^{\ast
}=\mathbf{I.}\label{eq:eigenvectors}%
\end{equation}
Equivalently, in the basis $\left\{  \vec{u}_{1},\ldots,\vec{u}_{n}\right\}
$, the covariance $\mathbf{FCF}^{\ast}$of $\mathbf{F}\vec{X}^{i}$ is diagonal,
with the diagonal elements $\lambda_{i}$. This is the situation, e.g., when
$\vec{X}^{i}$ are sampled from a second-order stationary random field on a
rectangular mesh, and $\vec{u}_{i}$ is the Fourier basis. In the EnKF, the
ensemble members after the first analysis cycle are not independent, because
the sample covariance in the analysis step ties them together, but they
converge to independent random vectors as the ensemble size $N\rightarrow
\infty$ \citep{LeGland-2011-LSA,Mandel-2011-CEK}. The following theorem shows
that the spectral diagonal approximation has smaller expected error than the
sample covariance, in Frobenius norm.

\begin{theorem}
[Error of the spectral diagonal approximation]\label{thm:diag-err} Let
$\vec{X}^{k}\sim N\left(  \bar{\vec{X}},\mathbf{C}\right)  $,
$k=1,\ldots,N$, be independent, and the transformation $\mathbf{F}$ satisfy
(\ref{eq:eigenvectors}). Then, the expected squared errors in the Frobenius
norm of the sample covariance $\mathbf{C}^{N}$ (\ref{eq:samplecov}) and its
spectral diagonal approximation $\mathbf{D}^{N}$ (\ref{eq:sdapprox}) are
\begin{align}
\operatorname{E}\left[  \Vert\mathbf{C}-\mathbf{C}^{N}\Vert_{\mathrm{F}}%
^{2}\right]   &  =\frac{2}{N-1}\sum_{i=1}^{n}\lambda_{i}^{2}+\frac{1}{N-1}%
\sum_{\substack{i,j=1\\i\neq j}}^{n}\lambda_{i}\lambda_{j}%
,\label{eq:sample-cov-expected-squared-error}\\
\operatorname{E}\left[  \Vert\mathbf{C}-\mathbf{D}^{N}\Vert_{\mathrm{F}}%
^{2}\right]   &  =\frac{2}{N-1}\sum_{i=1}^{n}\lambda_{i}^{2}.
\label{eq:sd-expected-squared-error}%
\end{align}

\end{theorem}

\textbf{Proof.} Without loss of generality, assume that $\bar{\vec{X}%
}=\vec{0}$. The Frobenius norm of a square matrix $\mathbf{A=}\left[  \vec
{a}_{1},\ldots,\vec{a}_{n}\right]  $ is unitarily invariant, $\left\Vert
\mathbf{FAF}^{\ast}\right\Vert _{\mathrm{F}}^{2}=\left\Vert \mathbf{A}%
\right\Vert _{\mathrm{F}}^{2}$, because $\left\Vert \mathbf{FA}\right\Vert
_{\mathrm{F}}^{2}=%
%TCIMACRO{\dsum \limits_{i=1}^{n}}%
%BeginExpansion
{\displaystyle\sum\limits_{i=1}^{n}}
%EndExpansion
\left\Vert \mathbf{Fa}_{i}\right\Vert ^{2}=%
%TCIMACRO{\dsum \limits_{i=1}^{n}}%
%BeginExpansion
{\displaystyle\sum\limits_{i=1}^{n}}
%EndExpansion
\left\Vert \vec{a}_{i}\right\Vert ^{2}=\left\Vert \mathbf{A}\right\Vert
_{\mathrm{F}}^{2}=\left\Vert \mathbf{A}^{\ast}\right\Vert _{\mathrm{F}}^{2}$.
Thus,
\[
\operatorname{E}\left[  \Vert\mathbf{C}-\mathbf{C}^{N}\Vert_{\mathrm{F}}%
^{2}\right]  =\operatorname{E}\left[  \Vert\mathbf{C}_{\mathbf{F}}%
-\mathbf{C}_{\mathbf{F}}^{N}\Vert_{\mathrm{F}}^{2}\right]  =\sum_{i,j=1}%
^{n}\operatorname{E}\left[  (\mathbf{C}_{\mathbf{F}})_{i,j}-(\mathbf{C}%
_{\mathbf{F}}^{N})_{i,j}\right]  ^{2}=\sum_{i,j=1}^{n}\operatorname{Var}%
\left[  (\mathbf{C}_{\mathbf{F}}^{N})_{i,j}\right]  ,
\]
because the sample covariance is unbiased, $\operatorname{E}\left[
(\mathbf{C}_{\mathbf{F}}^{N})_{i,j}\right]  =\left(  \mathbf{C_{F}}\right)
_{i,j}$. Lemma \ref{lem:CN_variance_of_elements} in the Appendix now gives
(\ref{eq:sample-cov-expected-squared-error}). To prove
(\ref{eq:sd-expected-squared-error}), we consider the diagonal entries in the
spectral domain,
\[
\operatorname{E}\left[  \Vert\mathbf{C}-\mathbf{D}^{N}\Vert_{\mathrm{F}}%
^{2}\right]  =\operatorname{E}\left[  \left\vert \mathbf{C}_{\mathbf{F}%
}-\mathbf{D}_{\mathbf{F}}^{N}\right\vert _{\mathbf{F}}^{2}\right]  =\sum
_{i=1}^{N}\operatorname{E}\left[  \left\vert \left(  \mathbf{C_{F}}\right)
_{i,i}-(\mathbf{C}_{\mathbf{F}}^{N})_{i,i}\right\vert ^{2}\right]  =\sum
_{i=1}^{n}\operatorname{Var}(\mathbf{C}_{F}^{N})_{i,i},
\]
and use Lemma \ref{lem:CN_variance_of_elements} again. \rule{0.5em}{0.5em}

Since the eigenvalues of covariance are always nonnegative, we have
$\lambda_{i}\lambda_{j}\geq0$, therefore the spectral diagonal covariance
decreases the expected squared error of sample covariance:
\[
\operatorname{E}\left[  \Vert\mathbf{C}-\mathbf{D}^{N}\Vert_{\mathrm{F}}%
^{2}\right]  \leq\operatorname{E}\left[  \Vert\mathbf{C}-\mathbf{C}^{N}%
\Vert_{\mathrm{F}}^{2}\right]  ,
\]
with equality only if all $\lambda_{i}\lambda_{j}=0$, $i\neq j$, that is, only
in the degenerate case when the exact covariance $\mathbf{C}$ has rank at most
one. To compare the error terms further, note that $\left(  \sum_{i=1}%
^{n}\lambda_{i}\right)  ^{2}=\sum_{i,j=1}^{n}\lambda_{i}\lambda_{j}%
=\sum_{i,j=1,i\neq j}^{n}\lambda_{i}\lambda_{j}+\sum_{i=1}^{n}\lambda_{i}^{2}%
$, which shows that the error of the sample covariance depends on the
$\ell^{1}$ norm of the eigenvalues sequence,
\[
\operatorname{E}\left[  \Vert\mathbf{C}-\mathbf{C}^{N}\Vert_{\mathrm{F}}%
^{2}\right]  =\frac{1}{N-1}\left(  \sum_{k=1}^{n}\lambda_{k}^{2}+\left(
\sum_{k=1}^{n}\lambda_{k}\right)  ^{2}\right)  =\frac{1}{N-1}\left(
\left\Vert \left\{  \lambda_{k}\right\}  _{k=1}^{n}\right\Vert _{\ell^{2}}%
^{2}+\left\Vert \left\{  \lambda_{k}\right\}  _{k=1}^{n}\right\Vert _{\ell
^{1}}^{2}\right)
\]
while the error of the spectral diagonal approximation depends only on the
$\ell^{2}$ norm,
\[
\operatorname{E}\left[  \Vert\mathbf{C}-\mathbf{D}^{N}\Vert_{\mathrm{F}}%
^{2}\right]  =\frac{2}{N-1}\left\Vert \left\{  \lambda_{k}\right\}  _{k=1}%
^{n}\right\Vert _{\ell^{2}}^{2},
\]
which is weaker as the state dimension $n\rightarrow\infty$. The improvement
depends on the rate of decay of the eigenvalues as the index $k\rightarrow
\infty$. Note that the eigenvalues of the covariance (if it exists) of a
random element in an infinitely dimensional Hilbert space must satisfy the
trace condition $\sum_{k=1}^{\infty}\lambda_{k}<\infty$, e.g.,
\cite{DaPrato-2006-IIA}. The eigenvalues of the covariance in many physical
systems obey a power law, $\lambda_{k}\approx k^{-\alpha}$ with $\alpha>1$,
e.g., \cite{Gaspari-1999-CCF}. Suppose that $\lambda_{k}=ck^{-\alpha}$ and
$n\rightarrow\infty$. Then,
\begin{align*}
\left\Vert \left\{  \lambda_{k}\right\}  _{k=1}^{n}\right\Vert _{\ell^{2}%
}^{2}  &  \rightarrow\sum_{k=1}^{\infty}k^{-2\alpha}\approx%
%TCIMACRO{\dint \limits_{1}^{\infty}}%
%BeginExpansion
{\displaystyle\int\limits_{1}^{\infty}}
%EndExpansion
x^{-2\alpha}dx=\frac{1}{2\alpha-1},\\
\left\Vert \left\{  \lambda_{k}\right\}  _{k=1}^{n}\right\Vert _{\ell^{1}%
}^{2}  &  \rightarrow\sum_{k=1}^{\infty}k^{-\alpha}\approx%
%TCIMACRO{\dint \limits_{1}^{\infty}}%
%BeginExpansion
{\displaystyle\int\limits_{1}^{\infty}}
%EndExpansion
x^{-\alpha}dx=\frac{1}{\alpha-1},
\end{align*}
which gives the error ratio $\operatorname{E}\left[  \Vert\mathbf{C}%
-\mathbf{D}^{N}\Vert_{\mathrm{F}}^{2}\right]  /\operatorname{E}\left[
\Vert\mathbf{C}-\mathbf{C}^{N}\Vert_{\mathrm{F}}^{2}\right]  \rightarrow0$ as
$\alpha\rightarrow1_{+}$. Other considerations of similar ratios can be found
in \citet{Furrer-2007-EHP}. Theorem~\ref{thm:diag-err} is related to but
different from the estimate in \citet[eq. (12)]{Furrer-2007-EHP}, which
applies to the case when the mean known exactly rather than the sample
covariance here. Also, the analysis in \cite{Furrer-2007-EHP} is in the
physical domain rather than in the spectral domain.

\section{Spectral EnKF algorithms}

We will show that the analysis step can be implemented very efficiently in
cases of practical interest. We drop the ensemble members index in all update
formulas to make them more readable. Note that when using all the following
formulas, it is necessary to perturb the observations.

\subsection{State consisting of only one variable, completely observed}

Assume that the state consists of one variable, e.g., $\vec{X}\in
\mathbb{R}^{n}$, and that we can observe the whole system state, i.e., the
observation function is the identity, $\mathbf{H}=\mathbf{I}$, and
observations are $\vec{Y}\in\mathbb{R}^{n}$. Assume also that the observation
noise covariance matrix is $c\mathbf{I}$, where $c>0$ is a constant. In this
special case, we can do the whole update in the spectral space, since it is
possible to transform the innovation to the spectral space, and the analysis
step (\ref{eq:sdenkfupdate}) becomes
\[
\vec{X}^{\mathrm{a}}=\vec{X}-\mathbf{F}^{\ast}\mathbf{D}_{\mathbf{F}}%
^{N}\left(  \mathbf{D}_{\mathbf{F}}^{N}+c\mathbf{I}\right)  ^{-1}%
\mathbf{F}\left(  \vec{X}-\vec{Y}\right)  .
\]

Note that the matrices $\mathbf{D}_{\mathbf{F}}^{N}$ and $\mathbf{D}%
_{\mathbf{F}}^{N}+c\mathbf{I}$ are diagonal, so any operation with them, such
as inversion or multiplication, is very cheap. The matrix $\mathbf{F}$ is
never formed explicitly. Rather, the multiplications of $\mathbf{F}$ and
$\mathbf{F}^{\ast}$ times a vector are implemented by the fast Fourier
transform (FFT) or discrete wavelet transform (DWT). This is the base case of
the FFT EnKF \citep{Mandel-2010-DDC,Mandel-2010-FFT} and the wavelet EnKF
\citep{Beezley-2011-WEK}, respectively.

\subsection{Multiple variables on the same grid, one variable completely
observed}

\label{sec:multiple}

In a typical model, such as numerical weather prediction, the state consist
usually of more than one variable. Assume the state consist of $m$ different
variables all based on the same grid of length $n$. Then each variable can be
transformed to the spectral space independently, and we have the state vector
$\vec{X}\in\mathbb{R}^{n\cdot m}$ and the transformation matrix in the block
form
\begin{equation}
\vec{X}=\left[
\begin{matrix}
\vec{X}_{1}\\
\vec{X}_{2}\\
\vdots\\
\vec{X}_{m}%
\end{matrix}
\right]  ,\quad\mathbf{F}=\left[
\begin{matrix}
\widetilde{\mathbf{F}} & \boldsymbol{0} & \cdots & \boldsymbol{0}\\
\boldsymbol{0} & \widetilde{\mathbf{F}} &  & \vdots\\
\vdots &  & \ddots & \boldsymbol{0}\\
\boldsymbol{0} & \cdots & \boldsymbol{0} & \widetilde{\mathbf{F}}%
\end{matrix}
\right]  , \label{eq:multivariate}%
\end{equation}
where each block $\vec{X}_{1}$ is a vector of length $n$ and
$\widetilde{\mathbf{F}}$ is $n$ by $n$ transformation matrix.

Assume also that the whole state of the first variable $\vec{X}_{1}$ is
observed, and again the covariance of observation error is $c\mathbf{I}$. In
this case, the observation operator is one by $m$ block matrix of the form
$\mathbf{H}=[%
\begin{matrix}
\mathbf{I} & \boldsymbol{0} & \cdots & \boldsymbol{0}%
\end{matrix}
]$. In the proposed method, we approximate the crosscovariancess between the
variables also by the diagonal of the sample covariance in spectral space,
$\mathbf{D}_{\mathbf{F}}^{N}=\left[
\begin{matrix}
\mathbf{D}_{i,j}^{N}%
\end{matrix}
\right]  _{i,j=1}^{m}$, where $\mathbf{D}_{i,j}$ is matrix containing only
diagonal elements from the sample covariance matrix between transformed
variables $\widetilde{\mathbf{F}}\vec{X}_{i}$ and $\widetilde{\mathbf{F}}%
\vec{X}_{j}$. With this notation, the analysis step (\ref{eq:sdenkfupdate})
becomes
\begin{equation}
\vec{X}^{\mathrm{a}}=\left[
\begin{matrix}
\vec{X}_{1}^{\mathrm{a}}\\
\vdots\\
\vec{X}_{m}^{\mathrm{a}}%
\end{matrix}
\right]  =\left[
\begin{matrix}
\vec{X}_{1}\\
\vdots\\
\vec{X}_{m}%
\end{matrix}
\right]  -\left[
\begin{matrix}
\widetilde{\mathbf{F}}^{\ast}\mathbf{D}_{1,1}^{N}\\
\vdots\\
\widetilde{\mathbf{F}}^{\ast}\mathbf{D}_{m,1}^{N}%
\end{matrix}
\right]  \left(  \mathbf{D}_{1,1}^{N}+c\mathbf{I}\right)  ^{-1}%
\widetilde{\mathbf{F}}\left(  \vec{X}_{1}-\vec{Y}\right)  .
\label{eq:same-grid-analysis}%
\end{equation}
Note that again the matrix to be inverted is diagonal and full-rank, and the
transformation $\widetilde{\mathbf{F}}$ is implemented by call to FFT\ or DWT,
so the operations are computationally very efficient. A related method using
interpolation and projection was proposed for the case when the model
variables are defined on non-matching grids \citep{Beezley-2011-WEK}.

\subsection{Multiple variables on the same grid, one variable observed at a
small number of points}

\label{sec:small-data}

This situation occurs, e.g., when assimilated observations are from discrete
stations. In this case, the observation matrix is $\mathbf{H}=\left[
\begin{matrix}
\mathbf{H}_{1} & \boldsymbol{0} & \cdots & \boldsymbol{0}%
\end{matrix}
\right]  $, where $\mathbf{H}_{1}$ has a small number of rows, one for each
data points, and $\vec{X}$ and $\mathbf{F}$ are the same as in
Eq.~(\ref{eq:multivariate}). We substitute the diagonal spectral approximation
into the analysis step (\ref{eq:sdenkfupdate}) directly, and
(\ref{eq:same-grid-analysis}) becomes
\begin{equation}
\vec{X}^{\mathrm{a}}=\left[
\begin{matrix}
\vec{X}_{1}\\
\vdots\\
\vec{X}_{m}%
\end{matrix}
\right]  -\left[
\begin{matrix}
\widetilde{\mathbf{F}}^{\ast}\mathbf{D}_{1,1}^{N}\\
\vdots\\
\widetilde{\mathbf{F}}^{\ast}\mathbf{D}_{m,1}^{N}%
\end{matrix}
\right]  \widetilde{\mathbf{F}}\left(  \mathbf{H}_{1}\widetilde{\mathbf{F}%
}^{\ast}\mathbf{D}_{1,1}^{N}\widetilde{\mathbf{F}}\mathbf{H}_{1}^{\ast
}+\mathbf{R}\right)  ^{-1}\widetilde{\mathbf{F}}\left(  \vec{X}_{1}-\vec
{Y}\right)  . \label{eq:small-data}%
\end{equation}

The solution of a system of linear equations with the matrix $\mathbf{H}%
_{1}\widetilde{\mathbf{F}}^{\ast}\mathbf{D}_{1,1}^{N}\widetilde{\mathbf{F}%
}\mathbf{H}_{1}^{\ast}+\mathbf{R}$ in (\ref{eq:small-data}) does not present a
problem, because its dimension is small by assumption, and
$\widetilde{\mathbf{F}}\mathbf{H}_{1}^{\ast}$ is easy to compute explicitly by
the action of FFT on the columns of $\mathbf{H}_{1}^{\ast}$. Note that in this
case, the data noise covariance $\mathbf{R}$ may be arbitrary.

\subsection{State consisting of more variables, one partly observed}

\label{sec:augmented}

Consider the situation when the number of observation points is too large for
the method of Sect.~\ref{sec:multiple} to be feasible, but only one variable
on a contiguous part of the mesh is observed. The typical example of this type
may be radar images, which cover typically only a part of domain of the
numerical weather prediction model.

Suppose that observations $\left(  \vec{Y}\right)  _{j}$ of the values of the
first variable $\left(  \vec{X}_{1}\right)  _{j}$ are available only for a
subset of indices $j\in M\subset\left\{  1,\ldots,m\right\}  $. Augment the
forecast state by an additional variable $\vec{X}_{0}$. For $j=1,\ldots,m$,
set $\left(  \vec{X}_{0}\right)  _{j}=\left(  \vec{X}_{1}\right)  _{j}$ if
$j\in M$, $\left(  \vec{X}_{0}\right)  _{j}=\left(  \vec{Y}\right)  _{j}=0$ if
$j\notin M$. We can now use the analysis update (\ref{eq:same-grid-analysis})
with the augmented state $\widetilde{\vec{X}}=\left(  \vec{X}_{0},\vec{X}%
_{1},\ldots,\vec{X}_{m}\right)  $ and observation $\widetilde{\vec{Y}}=\left(
\vec{Y},\vec{0},\ldots,\vec{0}\right)  $, to get the augmented analysis
$\widetilde{\vec{X}}^{\mathrm{a}}=\left(  \vec{X}_{0}^{\mathrm{a}},\vec{X}%
_{1}^{\mathrm{a}},\ldots,\vec{X}_{m}^{\mathrm{a}}\right)  $, and drop $\vec
{X}_{0}^{\mathrm{a}}$.

Note that the innovations to the original variables are propagated through the
spectral diagonal approximation of cross covariance between the original and
augmented variables. Since this covariance is not spatially homogeneous, a
Fourier basis will not be appropriate, and computational experiments in
Sect.~\ref{sec:comp} confirm that wavelets indeed perform better.

\section{Computational experiments}

\label{sec:comp}

In all experiments, we use the usual twin experiment approach. A run of the
model from one set of initial conditions is used to generate a sequence of
states, which plays the role of truth. Data values were obtained by applying
the observation operator to the truth; the data perturbation was done only for
ensemble members within the assimilation algorithm. \textbf{ }A second set of
initial conditions is used for data assimilation and for a free run, with no
data assimilation, for comparison. The error of the free run should be an
upper bound on the error of a reasonable data assimilation method.

We evaluate the filter by the root mean square error, $\operatorname{RMSE}%
=\bigl(\frac{1}{n}\sum_{i=1}^{n}\left(  \left(  \vec{X}\right)  _{i}-\left(
\bar{\vec{X}}^{\mathrm{a}}\right)  _{i}\right)  ^{2}\bigr)^{1/2}$, where
$\bar{\vec{X}}^{a}$ is the analysis ensemble mean, $\vec{X}$ is the true
state, and $n$ is the number of the grid points $x_{i}$. In the case when the
state consist of more than one variable, such as in the shallow water
equations, we evaluate the error of each variable independently. While the
purpose of a single analysis step is to balance the uncertainties of the state
and the data rather than minimalize the RMSE, the RMSE values over multiple
time steps are used to evaluate how well the data assimilation fulfills its
overall purpose to track the truth.

We evaluate the RMSE of the the standard EnKF, marked as EnKF in the legend of
the figures, and the spectral diagonal EnKF with the discrete sine transform,
discrete cosine transform, and the Coiflet 2,4 discrete wavelet transform
\citep{Daubechies-1992-TLW}, marked as DST, DCT, and DWT, respectively.

\subsection{Lorenz 96}

%f
\begin{figure}[t]%
\begin{tabular}
[c]{cc}%
\includegraphics[width=7cm]{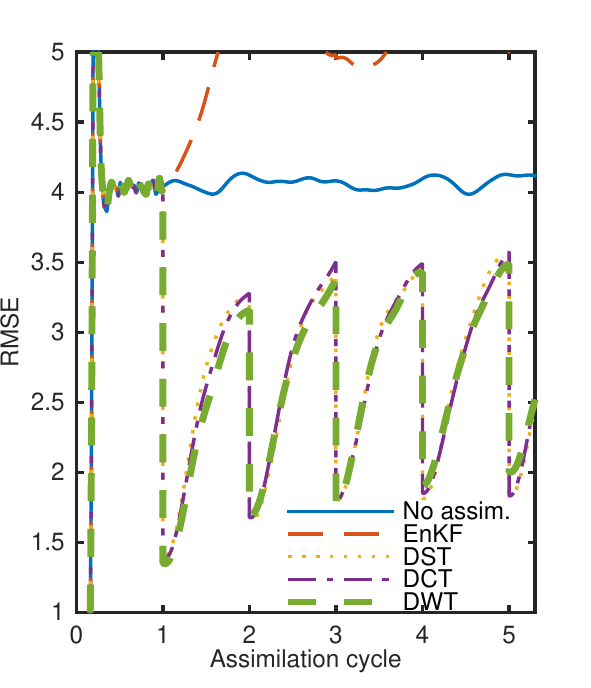} & \hspace{-0.3cm}
\includegraphics[width=7cm]{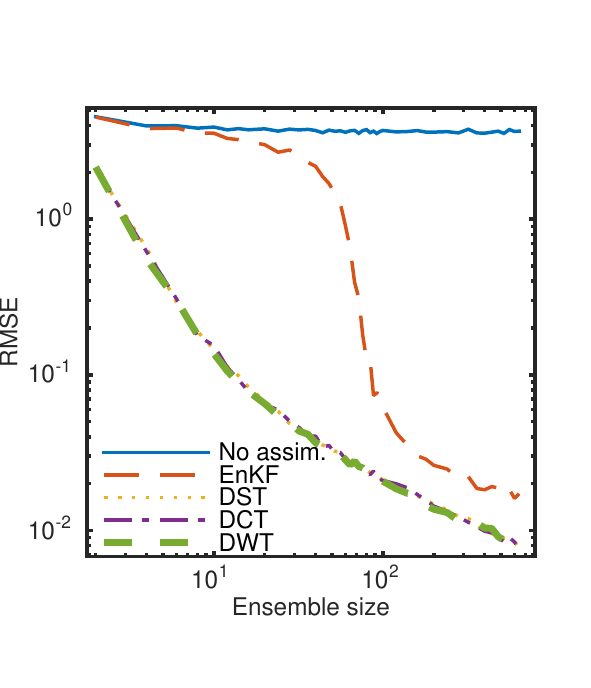}\\
(a) & (b)
\end{tabular}
%\includegraphics[width=12cm]{figure1.pdf}
%caption
\caption{Mean RMSE from 10 realizations for Lorenz 96 problem, the whole state
observed, state dimension 256 (a) increasing analysis cycles with ensemble
size 4, (b) increasing ensemble size, analysis cycle 1. }%
\label{fig:lorenz96}%
\end{figure}

%f
\begin{figure}[t]
\begin{tabular}
[c]{cc}%
\includegraphics[width=6.5cm]{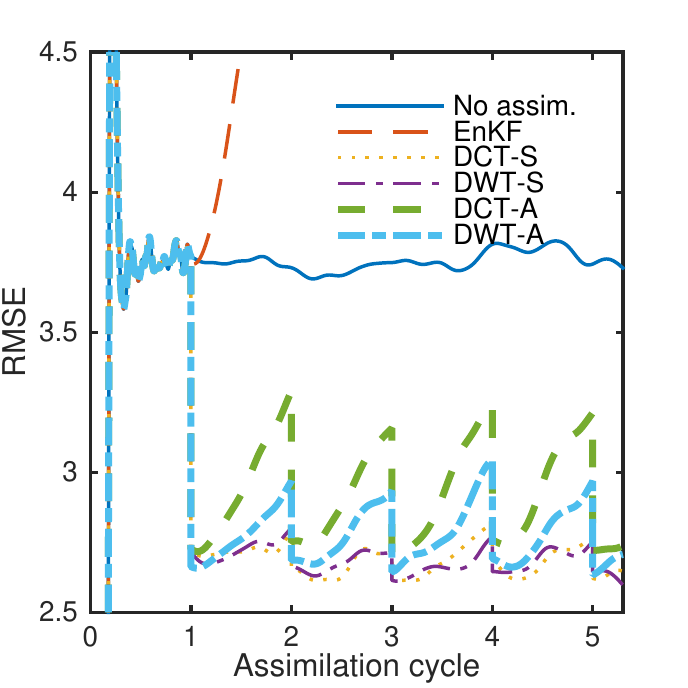} &
\hspace{-0.3cm}
\includegraphics[width=6.5cm]{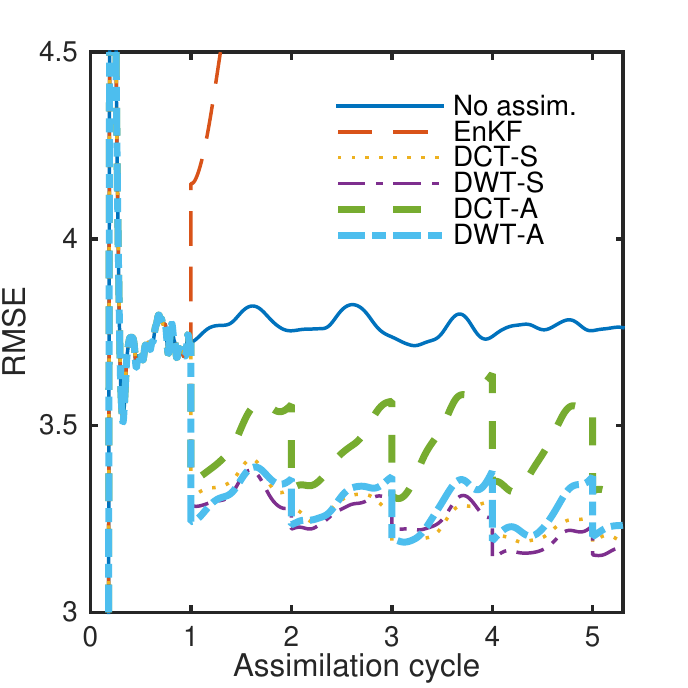}\\
(a) & (b)
\end{tabular}
%\includegraphics[width=12cm]{figure2.pdf}
%caption
\caption{Mean RMSE from 10 realizations for the Lorenz 96 problem, ensemble
size 16, state dimension 256. (a) first 128 points observed, (b) first 64
points observed. }%
\label{fig:lorenz96_EnsSize_16_ObsP}%
\end{figure}

In the Lorenz 96 model \citep{Lorenz-2006-PPP}, the state consists of one
variable $\vec{X}_{t}\in\mathbb{R}^{K}$, $\vec{X}_{t}=\left(  x_{1}%
,\ldots,x_{K}\right)  $, governed by the differential equations
\[
\frac{dx_{j}}{dt}=x_{j-1}x_{j+1}-x_{j-1}x_{j-2}-x_{j}+F,\quad j=1,\ldots,K,
\]
where the values of $x_{j-K}$ and $x_{j+K}$ are defined to be equal to $x_{j}$
for each $j=1,\ldots,K$, and $F$ is a parameter. We set the parameter $F=8$,
which causes the system to be strongly chaotic. The timestep of model was set
to $0.01%
%TCIMACRO{\TeXButton{unit}{\,\unit{s}}}%
%BeginExpansion
\,\unit{s}%
%EndExpansion
$ and the analysis cycle was $1%
%TCIMACRO{\TeXButton{unit}{\,\unit{s}}}%
%BeginExpansion
\,\unit{s}%
%EndExpansion
$. The data covariance was diagonal, with diagonal entries equal to $0.04$.
The ensemble and the initial conditions for the truth were generated by
sampling from $N(0.0005,0.01)$. The the ensemble and the truth were moved
forward for 10 second, then the assimilation starts\emph{.}

In the case when the whole state is observed, spectral filters with ensemble
size $N=4$ (Fig.~\ref{fig:lorenz96}a) already decrease the error significantly
compared to a run with no assimilation, while the standard EnKF actually
increases the error. For all filters, the error eventually decreases with the
ensemble size at the standard rate $N^{-1/2}$, but spectral EnKF shows the
error decrease from the start, while the EnKF\ lags until the ensemble size is
comparable to the state dimension, and even then its RMSE is significantly
higher (Fig.~\ref{fig:lorenz96}b).

Next, consider the case when only the first $m$ points of a grid are observed.
In the legend, DCT-S and DWT-S are the method with the discrete cosine
transform, and the Coiflet 2,4 discrete wavelet transform, respectively, with
the standard analysis update (\ref{eq:sdenkfupdate}), while DCT-A and DWT-A
use the augmented state method from Sect.~\ref{sec:augmented}. Figure~
\ref{fig:lorenz96_EnsSize_16_ObsP} shows that the spectral diagonal method
decrease the RMSE, while the standard EnKF is unstable. This observation is
consistent with the result of \cite{Kelly-2014-WAE}, which shows that, for a
class of dynamical systems, the EnKF remains within a bounded distance of
truth if sufficiently large covariance inflation is used and if the whole
state is observed. The augmented state method DWT-A with wavelet
transformation gave almost the same analysis error as DCT-S, which is using
the spectral diagonal filter with the exact observation matrix, while the
cosine basis, which implies a homogenenous random field, resulted in a much
larger error (method DCT-A). A similar behavior was seen with a smaller number
of observed points as well, but the error reduction in spectral diagonal EnKF
was smaller (not shown).

\subsection{Shallow water equations}

%f

The shallow water equations can serve as a simplified model of atmospheric
flow. The state $\vec{Y}=\left(  \vec{h},\vec{u},\vec{v}\right)  $ consists of
water level height $\vec{h}$ and momentum $\vec{u},\vec{v}$ in $x$ and $y$
directions, governed by the differential equations of conservation of mass and
momentum,
\begin{align*}
\frac{\partial\vec{h}}{\partial t}+\frac{\partial(\vec{uh})}{\partial x}%
+\frac{\partial(\vec{vh})}{\partial y}  &  =0,\\
\frac{\partial(\vec{h}\vec{u})}{\partial t}+\frac{\partial}{\partial x}\left(
\vec{h}\vec{u}^{2}+\frac{1}{2}g\vec{h}^{2}\right)  +\frac{\partial(\vec{h}%
\vec{uv})}{\partial y}  &  =0,\\
\frac{\partial(\vec{h}\vec{v})}{\partial t}+\frac{\partial(\vec{h}\vec{uv}%
)}{\partial x}+\frac{\partial}{\partial y}\left(  \vec{h}\vec{v}^{2}+\frac
{1}{2}g\vec{h}^{2}\right)   &  =0,
\end{align*}
where $g$ is gravity acceleration, with reflective boundary conditions, and
without Coriolis force or viscosity. The equations were discretized on a
rectangular grid size $64\times64$ with horizontal distance between grid
points $150%
%TCIMACRO{\TeXButton{km}{\,\unit{km}}}%
%BeginExpansion
\,\unit{km}%
%EndExpansion
$ and advanced by the Lax-Wendroff method with the time step $1%
%TCIMACRO{\TeXButton{s}{\,\unit{s}}}%
%BeginExpansion
\,\unit{s}%
%EndExpansion
$. The initial values where water level $\vec{h}=10%
%TCIMACRO{\TeXButton{km}{\,\unit{km}}}%
%BeginExpansion
\,\unit{km}%
%EndExpansion
$, plus Gaussian water raise of height $1%
%TCIMACRO{\TeXButton{km}{\,\unit{km}}}%
%BeginExpansion
\,\unit{km}%
%EndExpansion
$, width $32$ nodes, in the center of the domain, and $\vec{u}=\vec{v}=0$. See
\citet[Chapter 18]{Moler-2011-EWM} for details.

We have used two independent initial conditions, one used for the truth and
another for the ensemble and the free run. The only difference was the
location of the initial wave. Both states were moved forward for 4 hours. Then
the ensemble was created by adding random noise (with prescribed background
covariance). Then, all states were moved forward for another hour, and
assimilation starts $5%
%TCIMACRO{\TeXButton{h}{\,\unit{h}}}%
%BeginExpansion
\,\unit{h}%
%EndExpansion
$ after the model initialization. All assimilation methods start with the same
forecast in the first assimilation cycle.

\begin{figure}[t]%
\begin{tabular}
[c]{ccc}%
\includegraphics[width=4.7cm]{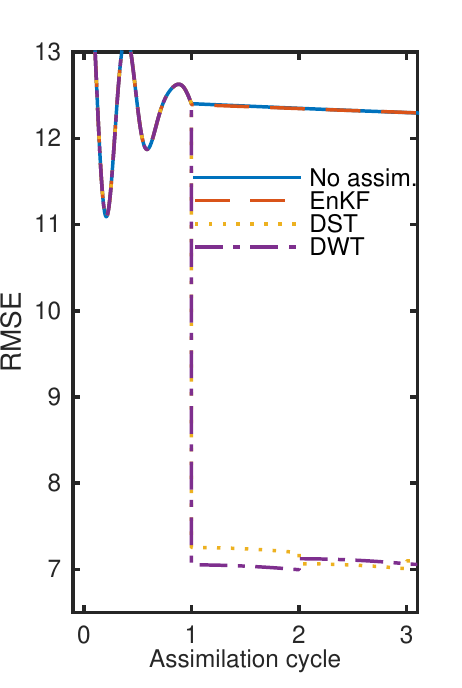} &
\hspace{-0.5cm}
\includegraphics[width=4.7cm]{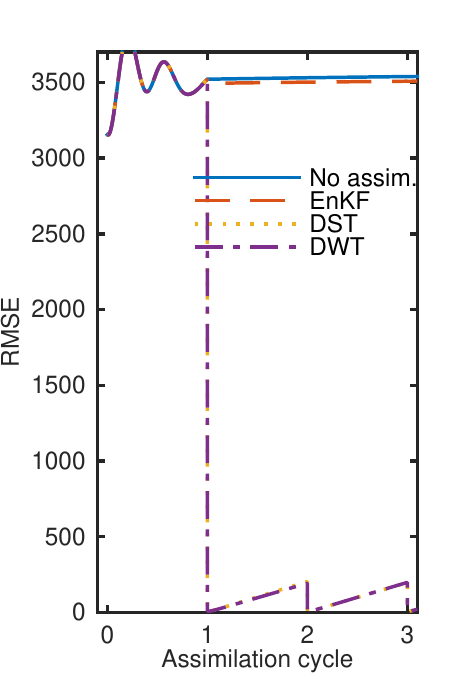} &
\hspace{-0.5cm}
\includegraphics[width=4.7cm]{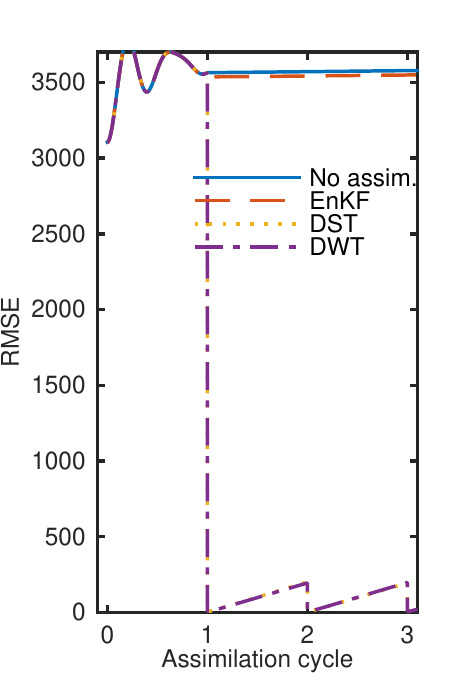}\\
(a) & (b) & (c)
\end{tabular}
%\includegraphics[width=12cm]{figure3.pdf}
%caption
\caption{RMSE of ensemble mean of one realization of three assimilation cycles
(f - forecast error, a - analysis error). Full state was observed. The length
of assimilation cycle 60 second, ensemble size 20. (a) water height (b)
momentum in the $x$ direction (c) momentum in the $y$ direction }%
\label{fig:waterwave_full_state_observations_rmse}%
\end{figure}

%f
\begin{figure}[t]
\begin{tabular}
[c]{ccc}%
\includegraphics[width=4.7cm]{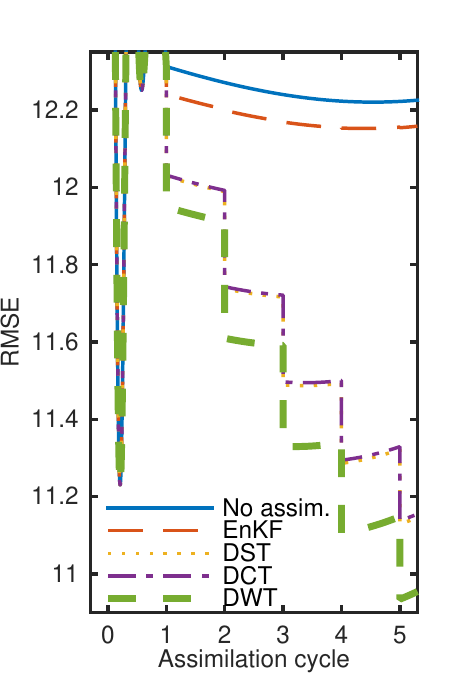} & \hspace{-0.5cm}
\includegraphics[width=4.7cm]{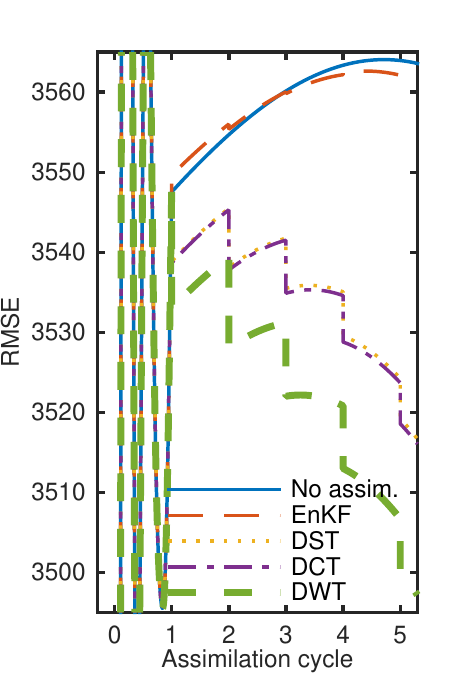} & \hspace{-0.5cm}
\includegraphics[width=4.7cm]{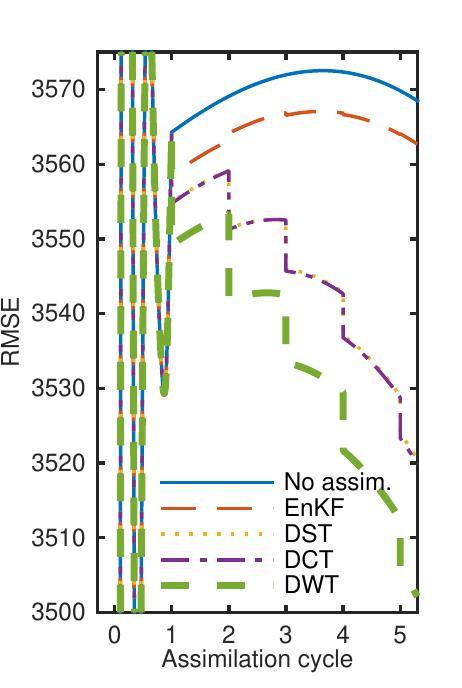}\\
(a) & (b) & (c)
\end{tabular}
%\includegraphics[width=12cm]{figure4.pdf}
%caption
\caption{Mean RMSE of ensemble mean from 5 independent repetitions. Ensemble
size 20, only water height observed. (a) water height (b) momentum in the $x$
direction (c) momentum in the $y$ direction }%
\label{fig:waterwave_1v}%
\end{figure}

The background covariance for initial ensemble perturbation was estimated
using samples taken every second from time $t_{\text{start}}=4%
%TCIMACRO{\TeXButton{h}{\,\unit{h}}}%
%BeginExpansion
\,\unit{h}%
%EndExpansion
$ to time $t_{\mathrm{end}}=6%
%TCIMACRO{\TeXButton{h}{\,\unit{h}}}%
%BeginExpansion
\,\unit{h}%
%EndExpansion
$, and modified by tapering the sample covariance matrix $\mathbf{C}_{N}$ as
$\mathbf{B}=\mathbf{C}_{N}\circ\mathbf{T},$ where the tapering matrix
$\mathbf{T}$ had the block structure
\[
\mathbf{T}=\left[
\begin{matrix}
\mathbf{A} & \mathbf{0} & \mathbf{0}\\
\mathbf{0} & \mathbf{A} & \mathbf{0}\\
\mathbf{0} & \mathbf{0} & \mathbf{A}%
\end{matrix}
\right]  +0.9\left[
\begin{matrix}
\mathbf{0} & \mathbf{A} & \mathbf{A}\\
\mathbf{A} & \mathbf{0} & \mathbf{A}\\
\mathbf{A} & \mathbf{A} & \mathbf{0}%
\end{matrix}
\right]  ,
\]
where the entry between nodes $\left(  i_{a},j_{a}\right)  $ and $\left(
i_{b},j_{b}\right)  $ is $A_{a,b}=\exp(-|i_{a}-i_{b}|)\exp(-|j_{a}-j_{b}|)$.
2D tensor product FFT and DWT were used in the diagonal spectral EnKF. The
observation error was taken with zero mean and variance $1000%
%TCIMACRO{\TeXButton{m^{2}}{\,\unit{m^{2}}}}%
%BeginExpansion
\,\unit{m^{2}}%
%EndExpansion
$ in $\vec{h}$ and $1000%
%TCIMACRO{\TeXButton{kg m s^{-1}}{\,\unit{kg}\,\unit{m}\,\unit{s}^{-1}}}%
%BeginExpansion
\,\unit{kg}\,\unit{m}\,\unit{s}^{-1}%
%EndExpansion
$ in $\vec{u}$ and $\vec{v}$. The forecast ensemble was created by adding
random noise with the covariance $\mathbf{B}$ $4%
%TCIMACRO{\TeXButton{h}{\,\unit{h}}}%
%BeginExpansion
\,\unit{h}%
%EndExpansion
$ after the model initialization. To relax the ensemble members, the model was
run for another hour before the assimilation started. So the first
assimilation was performed 5 hours after the model initialization. After the
first assimilation, another 4 assimilation cycles were performed every $60%
%TCIMACRO{\TeXButton{s}{\,\unit{s}}}%
%BeginExpansion
\,\unit{s}%
%EndExpansion
$.

When the full state is observed, the spectral diagonal method decreased the
RMSE in all variables dramatically
(Fig.~\ref{fig:waterwave_full_state_observations_rmse}), unlike the standard
EnKF. When only the water level is observed, the RMSE in spectral diagonal
EnKF decreases less, but still much more that in the standard EnKF
(Fig.~\ref{fig:waterwave_1v}).

%\conclusions
\section{Conclusions}

A version of the ensemble Kalman filter was presented, based on replacing the
sample covariance by its diagonal in the spectral space, which provides a
simple, efficient, and automatic localization. We have demonstrated efficient
implementations for several classes of observation operators and data
important in applications, including high-dimensional data defined on a
continuous part of the domain, such as radar or satellize images. The spectral
diagonal was proved rigorously to give a lower mean square error that the
sample covariance. Computational experimens with the Lorenz 96 problem and the
shallow water equations have shown that the method that the analysis error
drops very fast for small ensembles, and the method is stable over multiple
analysis cycles. The paper provides a new technology for data assimilation,
which can work with minimal computational resources, because an implementation
needs only an orthogonal transformation, such as the fast Fourier or discrete
wavelet transform, and manipulation of vectors and diagonal matrices.
Therefore, it should be of interest in applications.

%TCIMACRO{\TeXButton{Acknowledgements}{\begin{acknowledgements}
%This research was partially supported by the Czech Science
%Foundation under the grant GA13-34856S and  the U.S. National Science
%Foundation under the grant DMS-1216481.
%A part of this reseach was done when Ivan Kasanick\'{y} and Martin Vejmelka
%were visiting the University of Colorado Denver.
%\end{acknowledgements}}}%
%BeginExpansion
%\begin{acknowledgements}
\section*{Acknowledgements}
This research was partially supported by the Czech Science
Foundation under the grant GA13-34856S and  the U.S. National Science
Foundation under the grant DMS-1216481.
A part of this reseach was done when Ivan Kasanick\'{y} and Martin Vejmelka
were visiting the University of Colorado Denver.
%\end{acknowledgements}%
%EndExpansion

\bibliographystyle{copernicus}
\bibliography{../../references/other,../../references/geo,../../references/epi}

\newpage
\appendix

\section{Properties of sample covariance matrix}

Let $\vec{U}^{k}\sim N\left(  \vec{0},\mathbf{C}\right)  $ be independent
random vectors in $\mathbb{R}^{n}$ or $\mathbb{C}^{n}$. For each $\vec{U}^{k}%
$, we have the Karhunen-Lo\`{e}ve decomposition
\begin{equation}
\vec{U}^{k}=\sum_{j=1}^{n}\lambda_{j}^{1/2}\theta_{j,k}\vec{u}_{j},\quad
\theta_{j,k}\sim N(0,1)\text{ independent.} \label{eq:KL}%
\end{equation}
were $\lambda_{j}\geq0$ are the eigenvalues and $\vec{u}_{j}$ orthonormal
eigenvectors of the covariance matrix $\mathbf{C}$. Let $\mathbf{F}=\left[  \vec
{u}_{1},\ldots,\vec{u}_{n}\right]  ^{\ast}.$ By a direct computation, we have
in the basis of the eigenvectors:

\begin{lemma}
\label{lem:UF_cov} The random vector $\vec{U}_{\mathbf{F}}=\mathbf{FU}\sim
N\left(  \vec{0},\mathbf{C}_{\mathbf{F}}\right)  $, where $\mathbf{C}%
_{\mathbf{F}}=\mathbf{FCF}^{\ast}$ is a diagonal matrix with $\lambda
_{1},\ldots,\lambda_{n}$ on the diagonal.
\end{lemma}

Next, we use (\ref{eq:KL}) to compute an expansion of the sample covariance entries.

\begin{lemma}
\label{lem:CN_element} Let $\mathbf{C}_{\mathbf{F}}^{N}$ be the sample
covariance of $\vec{U}_{\mathbf{F}}^{1},\ldots,\vec{U}_{\mathbf{F}}^{N}$, cf.,
(\ref{eq:spectral-sample-cov}). Then,
\begin{equation}
(\mathbf{C}_{\mathbf{F}}^{N})_{i,j}=\frac{\left(  \lambda_{i}\lambda
_{j}\right)  ^{1/2}}{N-1}\left(  \sum_{k=1}^{N}\theta_{i,k}\theta_{j,k}%
-\frac{1}{N}\sum_{l=1}^{N}\theta_{i,l}\sum_{m=1}^{N}\theta_{j,m}\right)  .
\label{eq:CNF-ij}%
\end{equation}

\end{lemma}

\textbf{Proof.} From the definition of the sample covariance,
\begin{align*}
\left(  \mathbf{C}_{\mathbf{F}}^{N}\right)  _{i.j}  &  =\frac{1}{N-1}%
\sum_{k=1}^{N}\left(  \vec{U}_{\mathbf{F}}^{k}-\bar{\vec{U}}_{\mathbf{F}%
}\right)  _{i}\left(  \vec{U}_{\mathbf{F}}^{k}-\bar{\vec{U}}_{\mathbf{F}%
}\right)  _{j}^{\ast}\\
&  =\frac{1}{N-1}\sum_{k=1}^{N}\left(  \vec{U}_{\mathbf{F}}^{k}-\frac{1}%
{N}\sum_{l=1}^{N}\vec{U}_{\mathbf{F}}^{l}\right)  _{i}\left(  \vec
{U}_{\mathbf{F}}^{k}-\frac{1}{N}\sum_{m=1}^{N}\vec{U}_{\mathbf{F}}^{m}\right)
_{j}^{\ast}\\
&  =\frac{1}{N-1}\left(  \sum_{k=1}^{N}\left(  \vec{U}_{\mathbf{F}}%
^{k}\right)  _{i}\left(  \vec{U}_{\mathbf{F}}^{k\ast}\right)  _{j}-\frac{1}%
{N}\sum_{l=1}^{N}\left(  \vec{U}_{\mathbf{F}}^{k}\right)  _{l}\sum_{m=1}%
^{N}\left(  \vec{U}_{\mathbf{F}}^{l}\right)  _{m}\right) \\
&  =\frac{\left(  \lambda_{i}\lambda_{j}\right)  ^{1/2}}{N-1}\left(
\sum_{k=1}^{N}\theta_{i,k}\theta_{j,k}-\frac{1}{N}\sum_{l=1}^{N}\theta
_{i,l}\sum_{m=1}^{N}\theta_{j,m}\right)  .\quad\rule{0.5em}{0.5em}%
\end{align*}

Finally, we use the expansion (\ref{eq:CNF-ij}) to derive the variance of the
sample covariance entries.

\begin{lemma}
\label{lem:CN_variance_of_elements} The variance of each entry of
$\mathbf{C}_{\mathbf{F}}^{N}$ is
\[
\operatorname{Var}[(\mathbf{C}_{\mathbf{F}}^{N})_{i,j}]=\left\{
\begin{array}
[c]{ll}%
\frac{2\lambda_{i}^{2}}{N-1}\quad & \text{if }i=j,\\
\frac{\lambda_{i}\lambda_{j}}{N-1} & \text{if }i\neq j.
\end{array}
\right.
\]

\end{lemma}

\textbf{Proof.} The sample covariance is unbiased estimate of the true
covariance, so from Lemma \ref{lem:CN_element},
\begin{align}
\operatorname{Var}\left[  (\mathbf{C}_{\mathbf{F}}^{N})_{i,i}\right]   &
=\operatorname{E}\left[  (\mathbf{C}_{\mathbf{F}}^{N})_{i,i}-\operatorname{E}%
\left[  (\mathbf{C}_{\mathbf{F}}^{N})_{i,i}\right]  \right]  ^{2}%
=\operatorname{E}\left[  (\mathbf{C}_{\mathbf{F}}^{N})_{i,i}-(\mathbf{C}%
_{\mathbf{F}})_{i,i}\right]  ^{2}\nonumber\\
&  =\operatorname{E}\left[  \frac{\left(  \lambda_{i}\lambda_{i}\right)
^{1/2}}{N-1}\left(  \sum_{k=1}^{N}\theta_{i,k}^{2}-\frac{1}{N}\sum_{k,l=1}%
^{N}\left(  \theta_{i,k}\theta_{i,l}\right)  \right)  -\lambda_{i}\right]
^{2}\nonumber\\
&  =\frac{\lambda_{i}^{2}}{\left(  N-1\right)  ^{2}}\operatorname{E}\left[
\sum_{k=1}^{N}\theta_{i,k}^{2}\right]  ^{2}-\frac{2\lambda_{i}^{2}}{N\left(
N-1\right)  ^{2}}\operatorname{E}\left[  \sum_{k,l,m=1}^{N}\theta_{i,k}%
^{2}\theta_{i,l}\theta_{i,m}\right] \nonumber\\
&  +\frac{\lambda_{i}^{2}}{N^{2}\left(  N-1\right)  ^{2}}\operatorname{E}%
\left[  \sum_{k,l=1}^{N}\theta_{i,k}\theta_{i,l}\right]  ^{2}-\frac
{2\lambda_{i}^{2}}{\left(  N-1\right)  }\operatorname{E}\left[  \sum_{k=1}%
^{N}\theta_{i,k}^{2}\right] \nonumber\\
&  +\frac{2\lambda_{i}^{2}}{N\left(  N-1\right)  }\operatorname{E}\left[
\sum_{k,l=1}^{N}\theta_{i,k}\theta_{i,l}\right]  +\lambda_{i}^{2}.
\label{eq:CN_variation_diagonal_1}%
\end{align}
The random variables $\theta_{i,k}$ are i.i.d., so it follows that
\[
\operatorname{E}\left[  \theta_{i,k}\theta_{i,l}\theta_{i,m}\theta
_{i,n}\right]  =\left\{
\begin{array}
[c]{ll}%
3 & \quad\text{if }k=l=m=n,\\
1 & \quad\text{if }k=l,m=n,k\neq m,\\
1 & \quad\text{if }k=m,l=n,k\neq l,\\
1 & \quad\text{if }k=n,l=m,k\neq l,\\
0 & \quad\text{otherwise,}%
\end{array}
\right.
\]
and we can compute all the expected values in
Eq.~(\ref{eq:CN_variation_diagonal_1}),
\begin{align*}
\operatorname{E}\left[  \sum_{k=1}^{N}\theta_{i,k}^{2}\right]  ^{2}  &
=\sum_{k=1}^{N}\operatorname{E}\left[  \theta_{i,k}^{4}\right]  +\sum
_{k=1}^{N}\sum_{l=1,l\neq k}^{N}\operatorname{E}\left[  \theta_{i,l}^{2}%
\theta_{i,k}^{2}\right]  =3N+N(N-1)=N(N+2),\\
\operatorname{E}\left[  \sum_{k,l,m=1}^{N}\theta_{i,k}^{2}\theta_{i,l}%
\theta_{i,m}\right]   &  =\sum_{k=1}^{N}\operatorname{E}\left[  \theta
_{i,k}^{4}\right]  +\sum_{k,l=1,l\neq k}^{N}\operatorname{E}\left[
\theta_{i,k}^{2}\theta_{i,l}^{2}\right]  =3N+N(N-1)=N(N+2),\\
\operatorname{E}\left[  \sum_{k,l=1}^{N}\theta_{i,k}\theta_{i,l}\right]  ^{2}
&  =\sum_{k,l,m,n=1}^{N}\operatorname{E}\left[  \theta_{i,k}\theta_{i,l}%
\theta_{i,m}\theta_{i,n}\right]  =\sum_{k=1}^{N}\operatorname{E}\left[
\theta_{i,k}^{4}\right]  +3\sum_{k,l=1,l\neq k}^{N}\operatorname{E}\left[
\theta_{i,k}^{2}\theta_{i,l}^{2}\right]  =3N^{2},\\
\operatorname{E}\left[  \sum_{k=1}^{N}\theta_{i,k}^{2}\right]   &  =\sum
_{k=1}^{N}\operatorname{E}\left[  \theta_{i,k}^{2}\right]  =N,\\
\operatorname{E}\left[  \sum_{k,l=1}^{N}\theta_{i,k}\theta_{i,l}\right]   &
=\sum_{k=1}^{N}\operatorname{E}\left[  \theta_{i,k}^{2}\right]  =N.
\end{align*}
Together, we get
\[
\operatorname{Var}\left[  (\mathbf{C}_{\mathbf{F}}^{N})_{i,i}\right]
=\lambda_{i}^{2}\left(  \frac{N(N+2)}{(N-1)^{2}}-\frac{2(N+2)}{(N-1)^{2}%
}+\frac{3}{(N-1)^{2}}-\frac{2N}{N-1}+\frac{2}{N-1}+1\right)  =\frac
{2\lambda_{i}^{2}}{N-1}.
\]
The variance of the off-diagonal entry $(\mathbf{C}_{\mathbf{F}}^{N})_{i,j}$,
$i\neq j$, is
\begin{align}
\operatorname{Var}\left[  (\mathbf{C}_{\mathbf{F}}^{N})_{i,j}\right]   &
=\operatorname{E}\left[  (\mathbf{C}_{\mathbf{F}}^{N})_{i,j}-\operatorname{E}%
\left[  (\mathbf{C}_{\mathbf{F}}^{N})_{i,j}\right]  \right]  ^{2}%
=\operatorname{E}\left[  (\mathbf{C}_{\mathbf{F}}^{N})_{i,j}-(\mathbf{C}%
_{\mathbf{F}})_{i,j}\right]  ^{2}\nonumber\\
&  =\operatorname{E}\left[  \frac{\left(  \lambda_{i}\lambda_{j}\right)
^{1/2}}{N-1}\left(  \sum_{k=1}^{N}\theta_{i,k}\theta_{j,k}-\frac{1}{N}%
\sum_{k,l=1}^{N}\left(  \theta_{i,k}\theta_{j,l}\right)  \right)  -0\right]
^{2}\nonumber\\
&  =\frac{\lambda_{i}\lambda_{j}}{\left(  N-1\right)  ^{2}}\operatorname{E}%
\left[  \sum_{k=1}^{N}\theta_{i,k}\theta_{j,k}\right]  ^{2}-\frac{2\lambda
_{i}\lambda_{j}}{N\left(  N-1\right)  ^{2}}\operatorname{E}\left[
\sum_{k,l,m=1}^{N}\theta_{i,k}\theta_{j,k}\theta_{i,l}\theta_{j,m}\right]
\nonumber\\
&  +\frac{\lambda_{i}\lambda_{j}}{N^{2}\left(  N-1\right)  ^{2}}%
\operatorname{E}\left[  \sum_{k,l=1}^{N}\theta_{i,k}\theta_{j,l}\right]  ^{2}.
\label{eq:CN_variation_off-diagonal_1}%
\end{align}
The integrals in Eq.\ (\ref{eq:CN_variation_off-diagonal_1}) are
\begin{align*}
\operatorname{E}\left[  \sum_{k=1}^{N}\theta_{i,k}\theta_{j,k}\right]  ^{2}
&  =\sum_{k,l=1}^{N}\operatorname{E}\left[  \theta_{i,k}\theta_{j,k}%
\theta_{i,l}\theta_{j,l}\right]  =\sum_{k,l=1}^{N}\operatorname{E}\left[
\theta_{i,k}\theta_{i,l}\right]  \operatorname{E}\left[  \theta_{j,k}%
\theta_{j,l}\right] \\
&  =\sum_{k=1}^{N}\operatorname{E}\left[  \theta_{i,k}\theta_{i,k}\right]
\operatorname{E}\left[  \theta_{j,k}\theta_{j,l}\right]  =N,\\
\operatorname{E}\left[  \sum_{k,l,m=1}^{N}\theta_{i,k}\theta_{j,k}\theta
_{i,l}\theta_{j,m}\right]   &  =\sum_{k,l,m=1}^{N}\operatorname{E}\left[
\theta_{i,k}\theta_{i,l}\right]  \operatorname{E}\left[  \theta_{j,k}%
\theta_{j,m}\right]  =\sum_{k=1}^{N}\operatorname{E}\left[  \theta_{i,k}%
\theta_{i,k}\right]  \operatorname{E}\left[  \theta_{j,k}\theta_{j,k}\right]
=N,\\
\operatorname{E}\left[  \sum_{k,l=1}^{N}\theta_{i,k}\theta_{j,l}\right]  ^{2}
&  =\operatorname{E}\left[  \sum_{k=1}^{N}\theta_{i,k}\sum_{k,l=1}^{N}%
\theta_{j,l}\right]  ^{2}=\operatorname{E}\left[  \sum_{k=1}^{N}\theta
_{i,k}\right]  ^{2}\operatorname{E}\left[  \sum_{l=1}^{N}\theta_{j,l}\right]
^{2}=N^{2}.
\end{align*}
So, the variance of an off-diagonal element is $\operatorname{Var}\left[
(\mathbf{C}_{\mathbf{F}}^{N})_{i,j}\right]  =\frac{\lambda_{i}\lambda_{j}%
}{(N-1)^{2}}\left(  N-2+1\right)  =\frac{\lambda_{i}\lambda_{j}}{N-1}$%
.\quad\ \rule{0.5em}{0.5em}%

%FIGURES

\end{document}